\documentclass{emulateapj}
\usepackage{natbib}
\usepackage{rotating}
\usepackage{mathtools}
\usepackage{tablefootnote}
\usepackage{hyperref}

\newcommand{\ynyseed}{Y_\mathrm{n}/Y_\mathrm{seed}}
\newcommand{\yayseed}{Y_\alpha/Y_\mathrm{seed}}

\usepackage{graphicx}
\usepackage{amssymb}
\usepackage[usenames]{color}

\begin{document}

\title[Survey of astrophysical conditions in neutrino-driven supernova ejecta nucleosynthesis] {Survey of astrophysical conditions in neutrino-driven supernova ejecta nucleosynthesis}
\author{J. Bliss}
\affil{Institut f\"ur Kernphysik, Technische Universit\"at Darmstadt, Schlossgartenstr. 2,
Darmstadt 64289, Germany}
\email{julia.bliss@physik.tu-darmstadt.de}

\author{M. Witt}
\affil{Institut f\"ur Kernphysik, Technische Universit\"at Darmstadt, Schlossgartenstr. 2,
Darmstadt 64289, Germany}

\author{A. Arcones}
\affil{Institut f\"ur Kernphysik, Technische Universit\"at Darmstadt, Schlossgartenstr. 2,
Darmstadt 64289, Germany\\
\and GSI Helmholtzzentrum f\"ur Schwerionenforschung GmbH, Planckstr. 1, Darmstadt 64291, Germany}
\email{almudena.arcones@physik.tu-darmstadt.de}

\author{F. Montes}
\affil{National Superconducting Cyclotron Laboratory, Michigan
  State University, East Lansing, MI 48824, USA \\
\and Joint
  Institute for Nuclear Astrophysics, http://www.jinaweb.org}

\author{J. Pereira}
\affil{National Superconducting Cyclotron
  Laboratory, Michigan State University, East Lansing, MI 48824, USA \\
\and Joint Institute for Nuclear Astrophysics,
  http://www.jinaweb.org}

\begin{abstract} 

Core-collapse supernovae produce elements between Fe and Ag depending on the properties of the ejected matter.  Despite the fast progress in supernova simulations in the last decades, there are still uncertainties in the astrophysical conditions.  In this paper we investigate the impact of astrophysical uncertainties on the nucleosynthesis. Since a systematic study based on trajectories from hydrodynamic simulations is computationally very expensive, we rely on a steady-state model. By varying the mass and radius of the proto-neutron star as well as electron fraction in the steady-state model, we cover a wide range of astrophysical conditions. In our study, we find four abundance patterns which can be formed in neutron-rich neutrino-driven ejecta. This provides a unique template of trajectories that can be used to investigate the impact of nuclear physics input on the nucleosynthesis for representative astrophysical conditions.  Furthermore, we link these four patterns to the neutron-to-seed and alpha-to-seed ratios at $T=3$~GK. Therefore, our results give a good overview of the potential nucleosynthesis evolution which can occur in a supernova simulation.  

\end{abstract}
 \maketitle


\section{Introduction}
\label{sec:introduction}

Core-collapse supernovae represent the death of massive stars ($M\gtrsim 8M_\odot$), lead to the birth of neutron stars and stellar black holes, and they are the production site of many elements. They contribute to $1/3$ of the iron observed in our Galaxy, produce radioactive isotopes (e.g., $^{44}$Ti, $^{60}$Fe) whose decay has been observed \citep{Renaud.etal:2006,Grebenev:2012,Grefenstette.etal:2014,Wallner.etal:2016}, and synthesize heavy elements up to probably Ag/Cd \citep{Wanajo.etal:2011a,Wanajo.etal:2013b}. In some rare extreme cases where the explosion is driven by magnetic fields, even the heaviest elements may be produced by the r-process \citep{Winteler.etal:2012,Nishimura.etal:2015, Moesta.etal:2017,Halevi.Moesta:2018}.  The contribution of core-collapse supernovae to the chemical history of the universe needs to be studied based on self-consistent supernova simulations. This implies following the explosion and ejecta evolution for several seconds with three dimensional simulations in general relativity including detailed neutrino transport, and for several stellar progenitors. However, this is not possible today even if new efforts have been reported in this direction \citep{ Wanajo.etal:2011a, Wanajo.etal:2013a, Wanajo.etal:2013c, Wanajo.etal:2017, Harris.etal:2017, Eichler.etal:2017}. 

In this paper, we focus on the production of elements between iron and silver in the neutrino-driven ejecta. We follow a complementary approach to the expensive simulations by using a steady-state wind model which allows to study the neutrino-driven ejecta. The steady-state wind model has been proven to be very efficient in determining the required conditions for the r-process to occur in core-collapse supernovae \citep{Qian.Woosley:1996,Hoffman.etal:1997,Otsuki.etal:2000,Thompson.etal:2001,Wanajo.etal:2001}. We explore many combinations of electron fractions, neutron star masses and radii. These are input parameters for the wind equations and lead to a broad range of values for the wind parameters, namely entropy, expansion time scale, and electron fraction. Here, we investigate neutron-rich conditions and find a typical charged particle reaction process (sometimes also referred to as alpha process), and weak r-process nucleosynthesis. Current simulations predict proton-rich ejecta after the explosion (e.g., \citep{Bruenn.etal:2016}). However, uncertainties in neutrino-matter interactions may slightly change this \citep{MartinezPinedo.etal:2012,Roberts.etal:2012}. It has been found that there is also a small amount of neutron-rich matter that may still be ejected \citep{Wanajo.etal:2011a}. These ejecta are exposed only shortly to neutrinos and can be well described by our neutrino-driven wind model. Even if the amount of neutron-rich ejected matter is small, the contribution to the nucleosynthesis is very important because the mass fractions of elements heavier than iron are relatively high. In proton-rich conditions the ejected matter contains mainly alpha particles and protons, and therefore the mass fraction of heavy nuclei is very small \citep{Arcones.Bliss:2014, Arcones.Montes:2011}.

The paper is structure as followed. In Sect.~\ref{sec:method} the steady-state model and trajectories are described. We explain and compare the different nucleosynthesis groups created under different astrophysical conditions in Sect.~\ref{sec:results_nuc}. Finally, we summarize our results and conclude in Sect.~\ref{sec:conclusions}.

\section{Steady-state model and trajectories}
\label{sec:method}
We resort to steady-state models that were very successful in finding the appropriate conditions to produce the r-process in core-collapse supernovae~\citep{Qian.Woosley:1996,Hoffman.etal:1997,Cardall.Fuller:1997,Otsuki.etal:2000,Thompson.etal:2001,Wanajo.etal:2001}. With such a model, one can explore all possible conditions found in current and future simulations, as it was done for the r-process. Moreover, the trajectories obtained here mimic not only neutrino-driven wind ejecta, but also neutrino-driven ejecta in general, even if these are not supersonic winds. Therefore, our study can also roughly account for early neutrino-driven ejecta.

The steady-state model used here follows \cite{Otsuki.etal:2000} and it will be shortly summarized for completeness. Steady-state models rely on the fact that in the first few seconds after core-collapse, the proto-neutron star mass, radius, and (anti)neutrino luminosities and energies change slowly \citep{Qian.Woosley:1996} and time-dependencies can be neglected. We have compared the results of our steady-state model to simulations and found that, given the appropriate input parameters, it is possible to reproduce the evolution of the wind. However, in simulations there are also hydrodynamical features (like the reverse shock) that cannot be captured by a simple steady-state model \citep{Arcones.etal:2007,Arcones.Janka:2011}. In slightly neutron-rich winds, such hydrodynamical features have a small impact on the nucleosythesis in contrast to proton-rich conditions \cite{Wanajo.etal:2011b, Arcones.etal:2012, Arcones.Bliss:2014}.

The basic equations of the steady-state wind in a spherically symmetric Schwarzschild geometry are
\begin{gather}
 \dot{M} = 4\pi r^2 \rho\, v \, , \label{eq:ndw1} \\
 v\, \frac{dv}{dr} = -\, \frac{1}{\rho_{\mathrm{tot}}+P}\frac{dP}{dr}\left(1+v^2-\frac{2 M_\mathrm{ns}}{r}\right) - \frac{M_\mathrm{ns}}{r^2} \, , \label{eq:ndw2} \\
 \dot{q} = v\left(\frac{d\epsilon}{dr} - \frac{P}{\rho^2} \frac{d\rho}{dr}\right) \, , \label{eq:ndw3}
\end{gather}
where $\dot{M}$ is the constant mass outflow rate, $r$ is the distance from the center of the proto-neutron star, $\rho$ is the (baryon) mass density, $v$ is the radial velocity of the wind, $P$ the pressure, $\rho_{\mathrm{tot}} = \rho (1 + \epsilon)$ the total energy density with $\epsilon$ as the specific internal energy. Pressure and specific internal energy can be approximated as
\begin{eqnarray}
 P &=& \frac{11\pi^2}{180} T^4 + \frac{\rho}{m_\mathrm{N}} T \, , \label{eq:eos1} \\
 \epsilon &=& \frac{11\pi^2}{60} \frac{T^4}{\rho} + \frac{3}{2}\frac{T}{m_\mathrm{N}} \, , \label{eq:eos2}
\end{eqnarray}
assuming that matter is composed of non-relativistic nucleons, relativistic electrons and positrons, and photon radiation \citep{Otsuki.etal:2000}. The nucleon rest mass is \mbox{$m_\mathrm{N} = (m_\mathrm{p} + m_\mathrm{n})/2$}. Using these full set of equations, pressure, temperature, velocity, and density  can be derived as a function of the distance from the center of the proto-neutron star, given its star mass $M_\mathrm{ns}$, radius $R_\mathrm{ns}$, and neutrino and (anti)neutrino luminosities and energies.

The net heating rate from neutrino interactions with matter, $\dot{q}$, takes into account neutrino and antineutrino absorption on nucleons, electron and positron capture on nucleons, neutrino and antineutrino scattering off electrons and positrons, neutrino- antineutrino annihilation into electron and positron and its inverse (for more details see Eqs.~(8)-(16) of \cite{Otsuki.etal:2000}). These reactions depend on luminosities and energies for electron neutrino and antineutrino and on a third neutrino flavour that accounts for muon and tau neutrinos and antineutrinos. These neutrino quantities are all input parameters in the steady-state model. Since varying all of them is too expensive, we use the electron fraction to constrain them. We assume $\dot Y_{\mathrm{e}} = 0$, electron/positron capture negligible, and an initial composition consisting mainly of neutrons and protons. Then, the $Y_{\mathrm{e}}$ follows:
\begin{equation}
Y_{\mathrm{e}} = \left[ 1 + \frac{L^n_{\bar{\nu}_{\mathrm{e}}} \langle\sigma_{\bar{\nu}_{\mathrm{e}}p}\rangle}{L^n_{\nu_{\mathrm{e}}} \langle\sigma_{\nu_{\mathrm{e}}n}\rangle}\right]^{-1}, \label{eq:ye}
\end{equation}
where $L_{\nu}^n=L_{\nu} / \langle E_{\nu}\rangle$ is the number luminosity and is assumed to be the same for electron neutrinos and antineutrinos.  The electron neutrino energy luminosity and energy are kept constant ($\langle E_{\nu_{\mathrm{e}}}\rangle =16.66$~MeV and $L_{\nu_{\mathrm{e}}}= 2 \cdot 10^{51}$~ergs/s \citep{Arcones.etal:2007}). The cross sections for electron neutrino absorption on neutrons ($\langle\sigma_{\nu_{\mathrm{e}}n}\rangle$) and electron antineutrino absorption on protons ($\langle\sigma_{\bar{\nu}_{\mathrm{e}}p}\rangle$) depend on the neutrino and antineutrino energies. Therefore, for a fixed $\langle E_{\nu_{\mathrm{e}}}\rangle$ and a given $Y_{\mathrm{e}}$, one can calculate the antineutrino energy from Eq.~\ref{eq:ye}. With this $\langle E_{\bar{\nu}_{\mathrm{e}}}\rangle$ and the assumption of equal number luminosities, $L_{\bar{\nu}_{\mathrm{e}}}$ is fixed. The electron fraction is the main nucleosynthesis parameter because it determines the initial composition. For given $Y_{\mathrm{e}}$, the electron neutrino energy and luminosity have a small impact on the abundances due to the formation of alpha particle that are not considered in Eq.~\ref{eq:ye}. Therefore, keeping the electron neutrino energy and luminosity constant is justified and allows us to use the electron fraction as input parameter. 


The solutions of Eqs.~\eqref{eq:ndw1}--\eqref{eq:ndw3} depend on the mass outflow rate \citep{Duncan.etal:1986}. For instance, for large enough mass outflow ($\dot{M}= \dot{M}_\mathrm{crit}$), the velocity reaches the speed of sound corresponding to the \emph{wind} (or supersonic) solution. The so-called \emph{breeze} (or subsonic) solutions are found for $\dot{M} < \dot{M}_{\mathrm{crit}}$. If $\dot{M} > \dot{M}_{\mathrm{crit}}$, one gets unphysical solutions where the mass outflow experiences an infinite acceleration. $\dot{M}_\mathrm{crit}$ depends on the neutron star and neutrino properties.

We vary the input of the steady-state equations to cover all possible conditions of the neutrino-driven ejecta. The range of neutron star masses and radii have been chosen taking into account current observational and theoretical constraints for neutron stars and neutron matter (see e.g., \cite{Lattimer.etal:2016}). The values for the input quantities are given in Tab.~\ref{tab:wind_input} together with the values from \cite{Otsuki.etal:2000} and \cite{Thompson.etal:2001} for comparison. Here, we have focussed in neutron-rich conditions because we want to explore the weak r-process and charged particle reactions. By changing (anti)neutrino luminosities, energies, and $Y_{\mathrm{e}}$, one can also investigate proton-rich conditions. Note that in Tab.~\ref{tab:wind_input} our values partially overlap  with those of \cite{Otsuki.etal:2000} and \cite{Thompson.etal:2001}, this implies that we also find some extreme cases that produce r-process. However, we do not consider such extreme trajectories because their conditions are inconsistent with current supernova models. 

\begin{table}[!htb]
 \begin{center}
  \caption{Comparison between input parameters in the steady-state models used in this study, \cite{Otsuki.etal:2000} and \cite{Thompson.etal:2001}.}
  \label{tab:wind_input}
  \vspace*{3mm}
  \begin{tabular}{l|ccc}
  \hline
  \hline
  &This work & Otsuki & Thompson \\
  \hline
 $M_{\mathrm{ns}}/M_{\odot}$&  $0.8 - 2$&  $1.2 - 2 $ & $1.4 - 2$\\
 $R_{\mathrm{ns}}/\mathrm{km}$& $9 - 30$ & $10$ & $10 - 20.3$\\
$Y_{\mathrm{e}}$& $0.4 - 0.49$ & $0.43 - 0.46$ & $0.45 - 0.495$\\
  \hline
  \end{tabular}
 \end{center}
\end{table}

\begin{figure}[h!]
\centering
\includegraphics[width=0.99\linewidth]{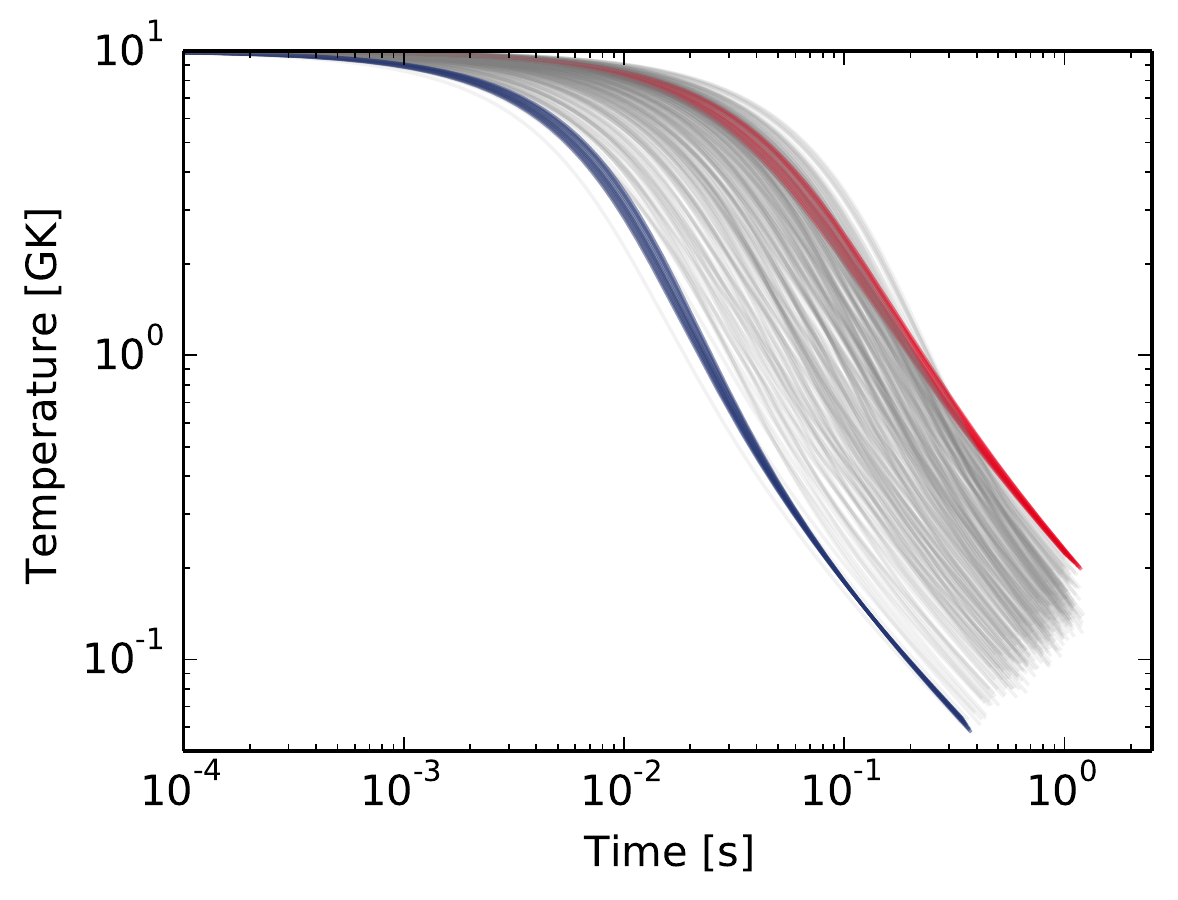}\\
\includegraphics[width=0.99\linewidth]{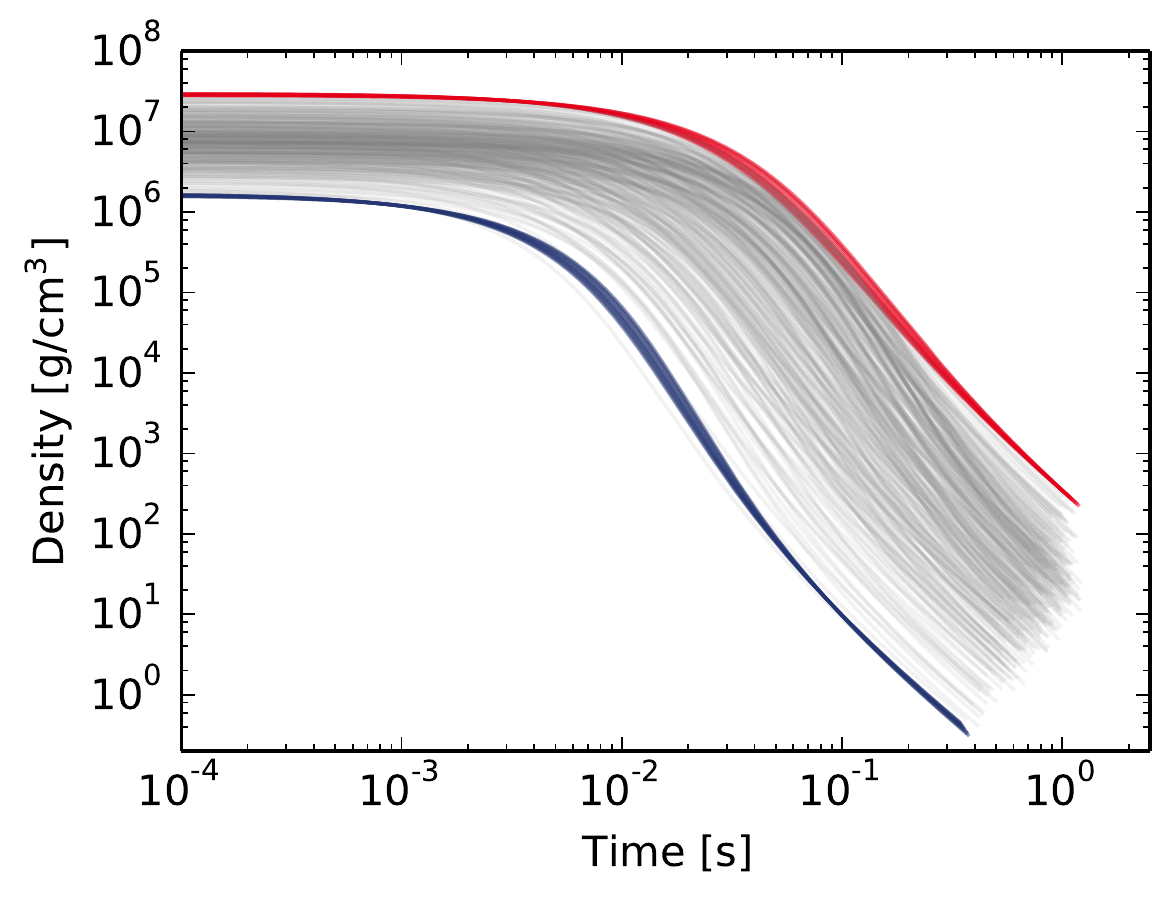}
\caption{Overview of temperature (top panel) and density evolution (bottom panel) of the steady-state trajectories
included in the present study (grey lines). Extreme trajectories
calculated with $R_{\mathrm{ns}}=30$~km, $M_{\mathrm{ns}}=0.8$~$M_{\odot}$  and $R_{\mathrm{ns}}=9.0$~km, $M_{\mathrm{ns}}=2.0$~$M_{\odot}$ are shown by the red and blue bands, respectively. The spread of the red and blue bands is due to the different electron fractions ($0.40 \leq Y_{\mathrm{e}} \leq 0.49$).}
\label{Fig.:Impact_MnsRnsLv_TempDens}
\end{figure}

The evolution of wind temperature and density as a function of time (after converting velocity as a function of the wind radius) is shown in Fig.~\ref{Fig.:Impact_MnsRnsLv_TempDens} for different combinations of $M_{\mathrm{ns}}$,  $R_{\mathrm{ns}}$, and $Y_{\mathrm{e}}$ (see Tab.~\ref{tab:wind_input}).  The most compact proto-neutron star ($M_\mathrm{ns} = 2M_{\odot}$ and $R_\mathrm{ns}=9\mathrm{km}$) results in a faster drop of the temperature and density. The highest temperatures and densities are obtained for the largest proto-neutron star radius and lowest proto-neutron star mass.  The width of each band is due to the variation of the electron fraction.

\begin{figure}[h!]
  \centering
  \includegraphics[width=1.\linewidth]{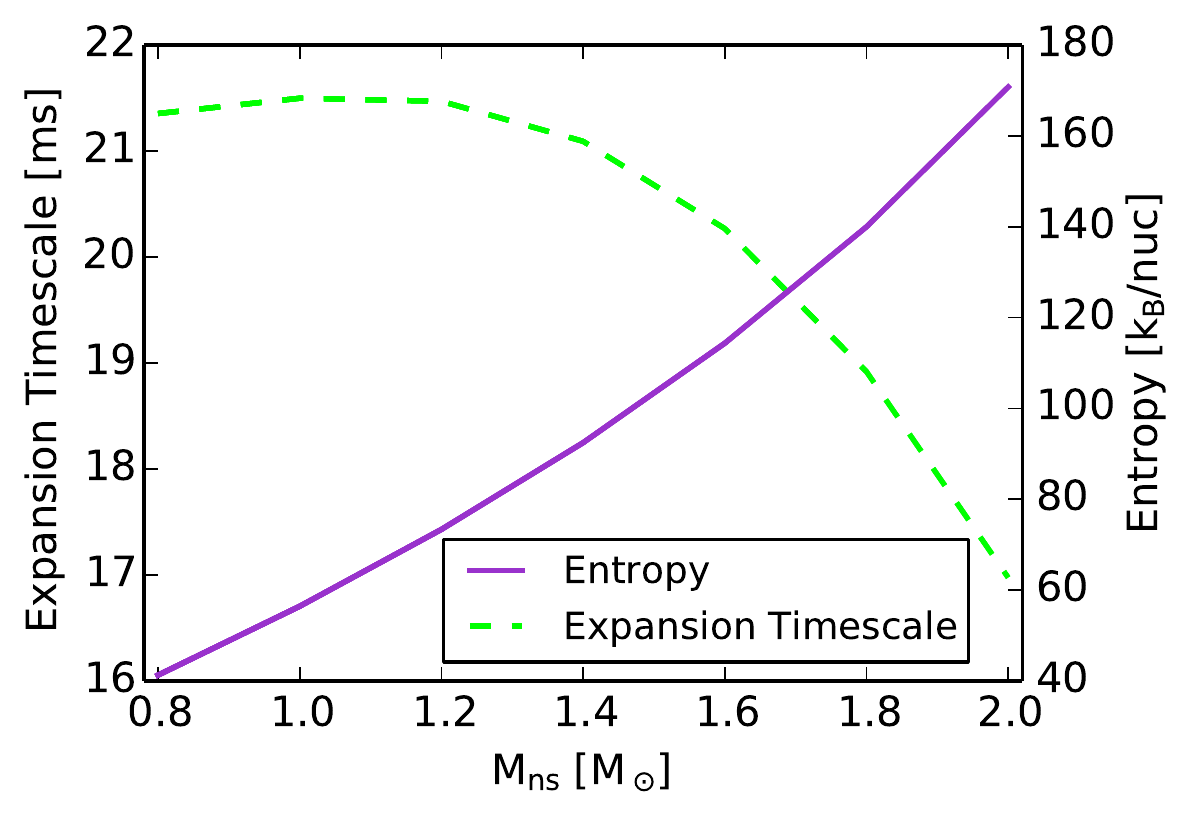} \\
  \includegraphics[width=1.\linewidth]{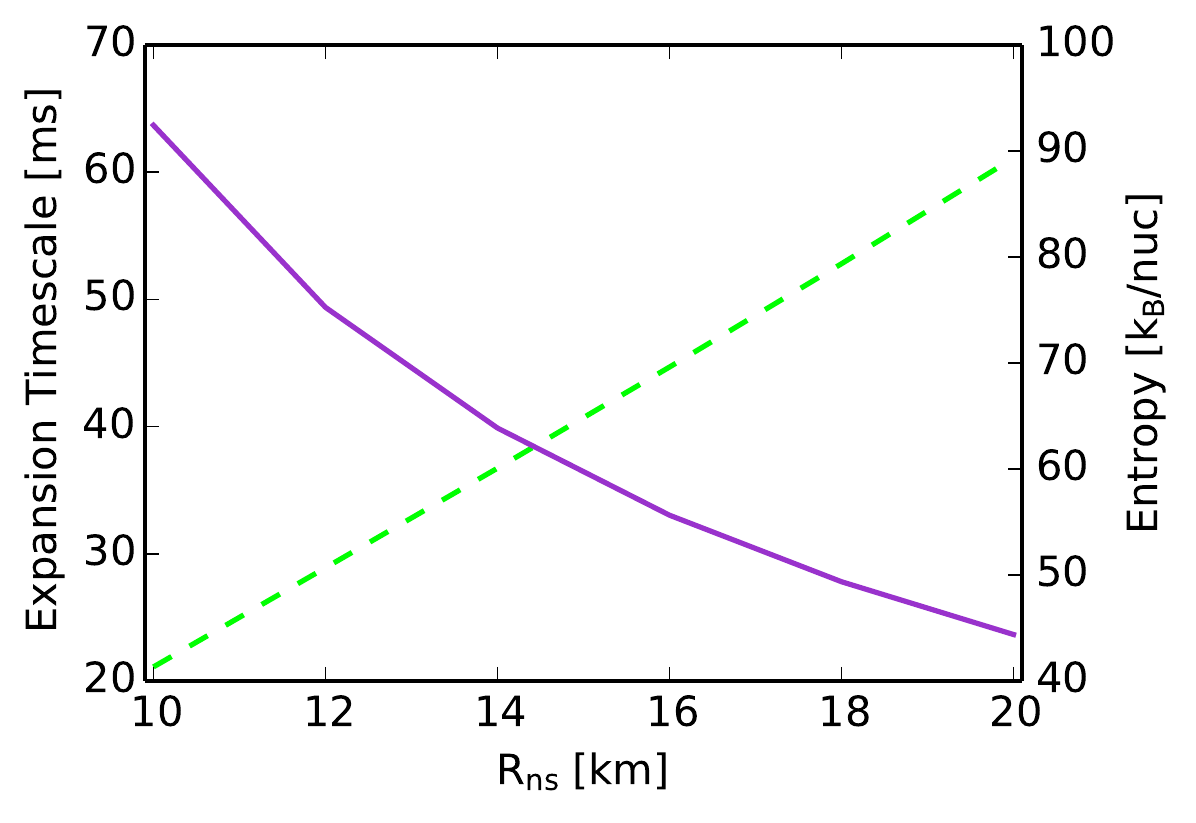}
\caption{Impact of the proto-neutron star mass ($M_{\mathrm{ns}}$) and radius ($R_{\mathrm{ns}}$) on the entropy (solid lines) and expansion time scale (dashed lines). The electron fraction is constant $Y_{\mathrm{e}}=0.45$. In the upper panel the proto-neutron star radius is kept constant and equal to 10~km, in the bottom panel the proto-neutron star mass is constant and equal to 1.4~$M_\odot$} \label{Fig.:Impact_MnsRnsLv_EntropyTimescale}
\end{figure}

Figure~\ref{Fig.:Impact_MnsRnsLv_EntropyTimescale} illustrates the dependance of the entropy \mbox{($S \propto T^{3}/\rho$)} and expansion time scale (defined as \mbox{$\tau = \left. r/v \right|_{T=0.5 \mathrm{MeV} \approx \mathrm{5 GK}}$} \citep{Qian.Woosley:1996}) on $M_\mathrm{ns}$ and $R_\mathrm{ns}$ assuming \mbox{$Y_{\mathrm{e}}=0.45$}. We chose a reference case, i.e., $M_{\mathrm{ns}}=1.4$~$M_{\odot}$ and $R_{\mathrm{ns}}=10$~km. As already explained in many wind studies (e.g., ~\cite{Cardall.Fuller:1997,Otsuki.etal:2000,Thompson.etal:2001,Wanajo.etal:2001}),  the wind entropy increases and the expansion time scale decreases as the proto-neutron star mass increases. Moreover, larger proto-neutron star radii lead to smaller entropies and longer expansion time scales. Therefore, a more compact proto-neutron star (i.e., more mass and/or smaller radius) ejects slightly less material due to the larger binding ($M/R$). In such a case, entropies are higher and  expansion time scales shorter due to the larger neutrino energy deposition that is necessary to unbound matter \citep{Cardall.Fuller:1997,Wanajo.etal:2001}.

\section{Characteristic nucleosynthesis patterns}
\label{sec:results_nuc}

We have calculated the nucleosynthesis for 2696 steady-state trajectories using the WinNET reaction network \citep{Winteler:2012,Winteler.etal:2012}. In the network, we consider 4412 nuclei from H to Ir including neutron- and proton-rich nuclei as well as stable ones. The reaction rates are taken from the JINA ReaclibV2.0 \citep{Cyburt.etal:2010} library. We use the same theoretical weak interaction rates and neutrino reactions on nucleons as in Ref.~\cite{Froehlich.etal:2006}. We start the calculation of every nucleosynthesis trajectory at 10~GK and assume nuclear statistical equilibrium (NSE) down to 8~GK. Weak reactions are not in equilibrium and thus we calculate their impact on $Y_{\mathrm{e}}$ during the whole evolution. At early times when the temperature is still high, matter is  close to the proto-neutron star and it consists mainly of neutrons and protons (photons dissociate any nuclei that forms). As matter expands and temperature decreases, alpha particles form and later these combine producing seed nuclei\footnote{Here, the seed abundance $Y_{\mathrm{seed}}$ is defined as the sum of the abundances of all nuclei heavier than helium.}. The subsequent evolution strongly depends on entropy, expansion time scale, and $Y_{\mathrm{e}}$. 

\begin{figure*}[h!]
\centering
\includegraphics[width=0.9\linewidth]{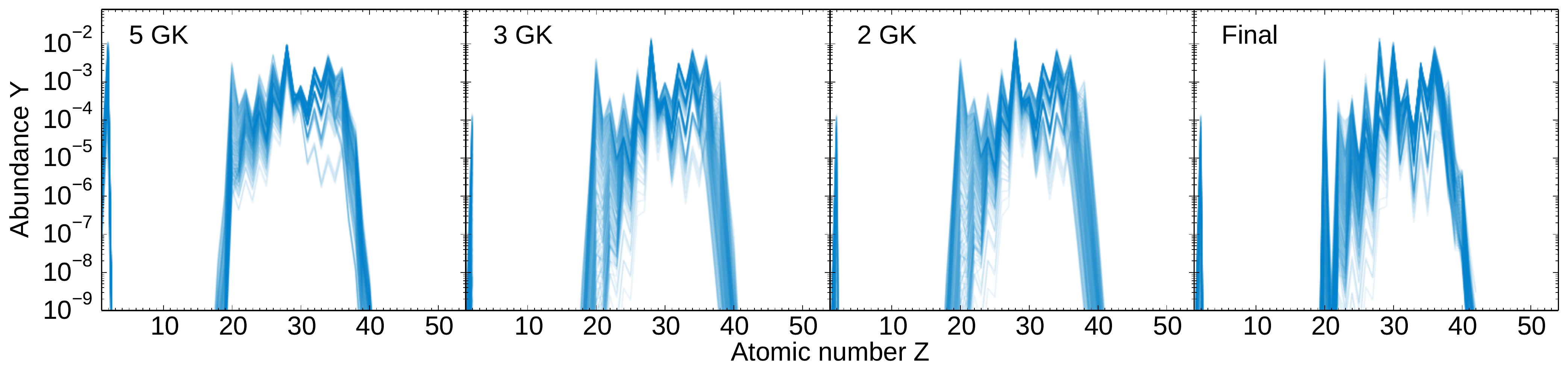}\\
\includegraphics[width=0.9\linewidth]{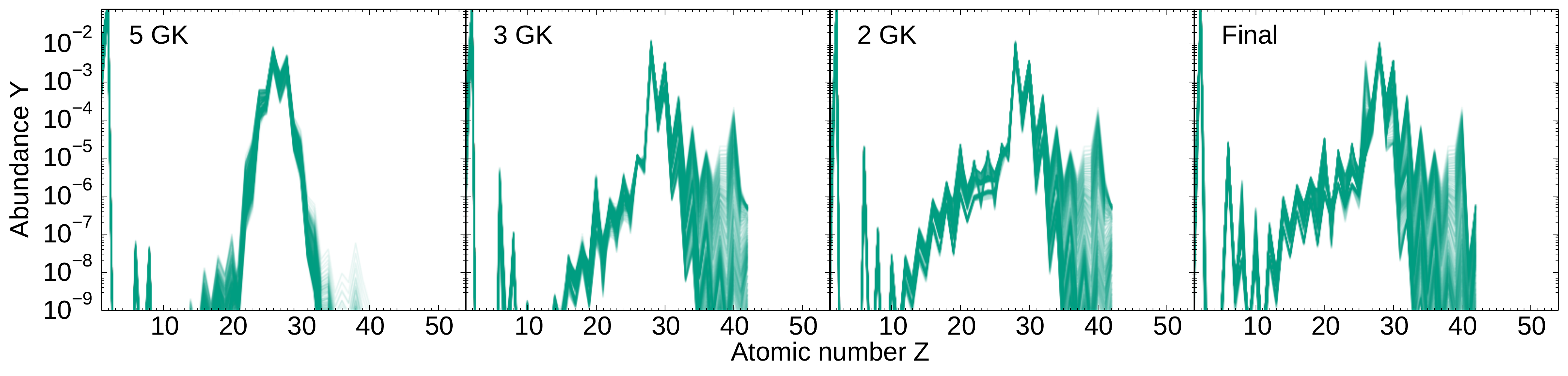}\\
\includegraphics[width=0.9\linewidth]{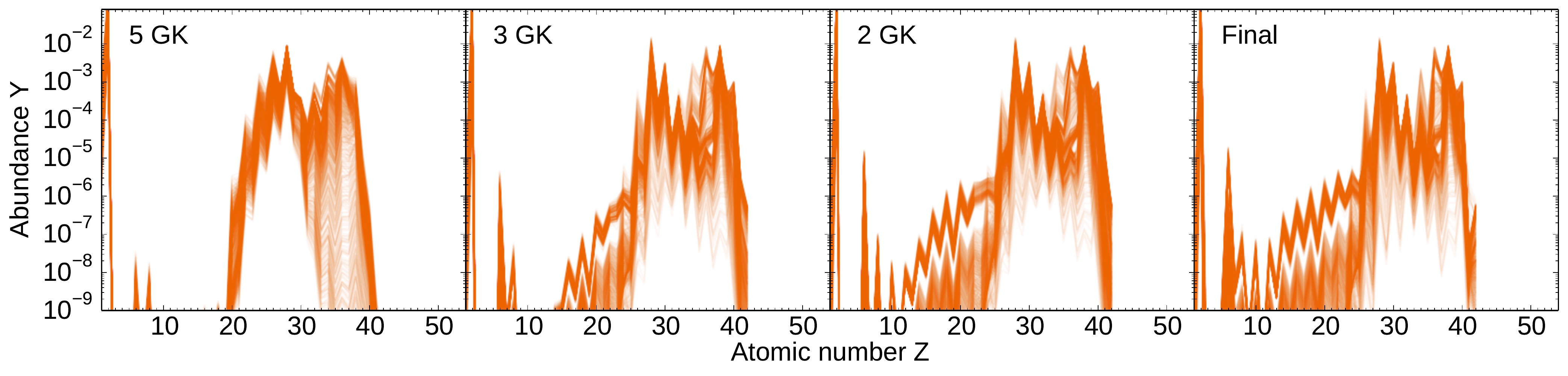}\\
\includegraphics[width=0.9\linewidth]{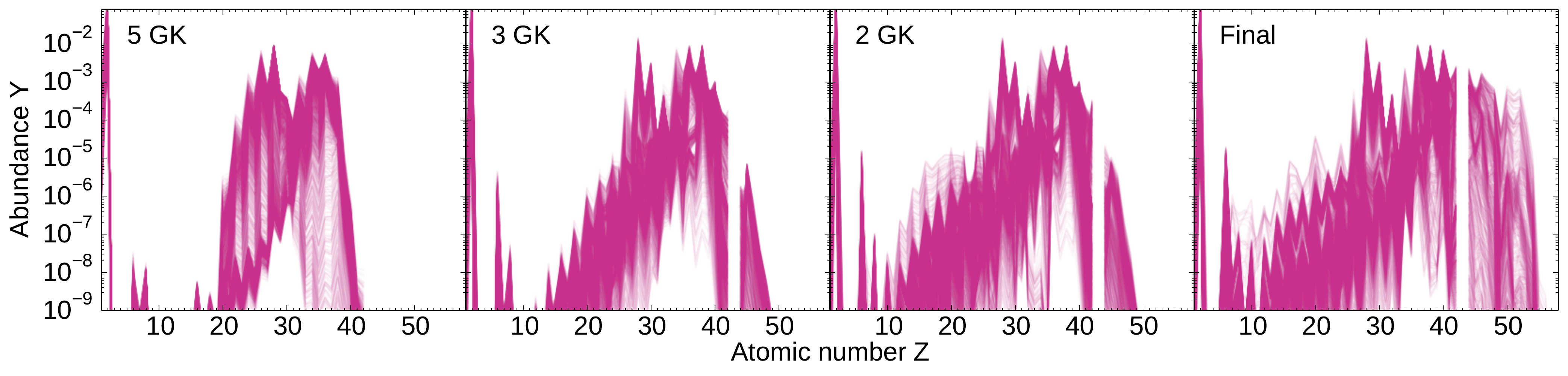}
\caption{Abundances of the nucleosynthesis groups NSE1 (first row), NSE2 (second row), CPR1 (third row), CPR2 (fourth row) at $T = 5, 3, 2$~GK, and after final decay. We apply the same color code to distinguish between the different groups as in Fig.~\ref{fig:YnYseedvsYalphaYseed}.}
\label{fig:nuc_types}
\end{figure*}

\begin{figure}[t!]
\centering
\includegraphics[width=1.\linewidth]{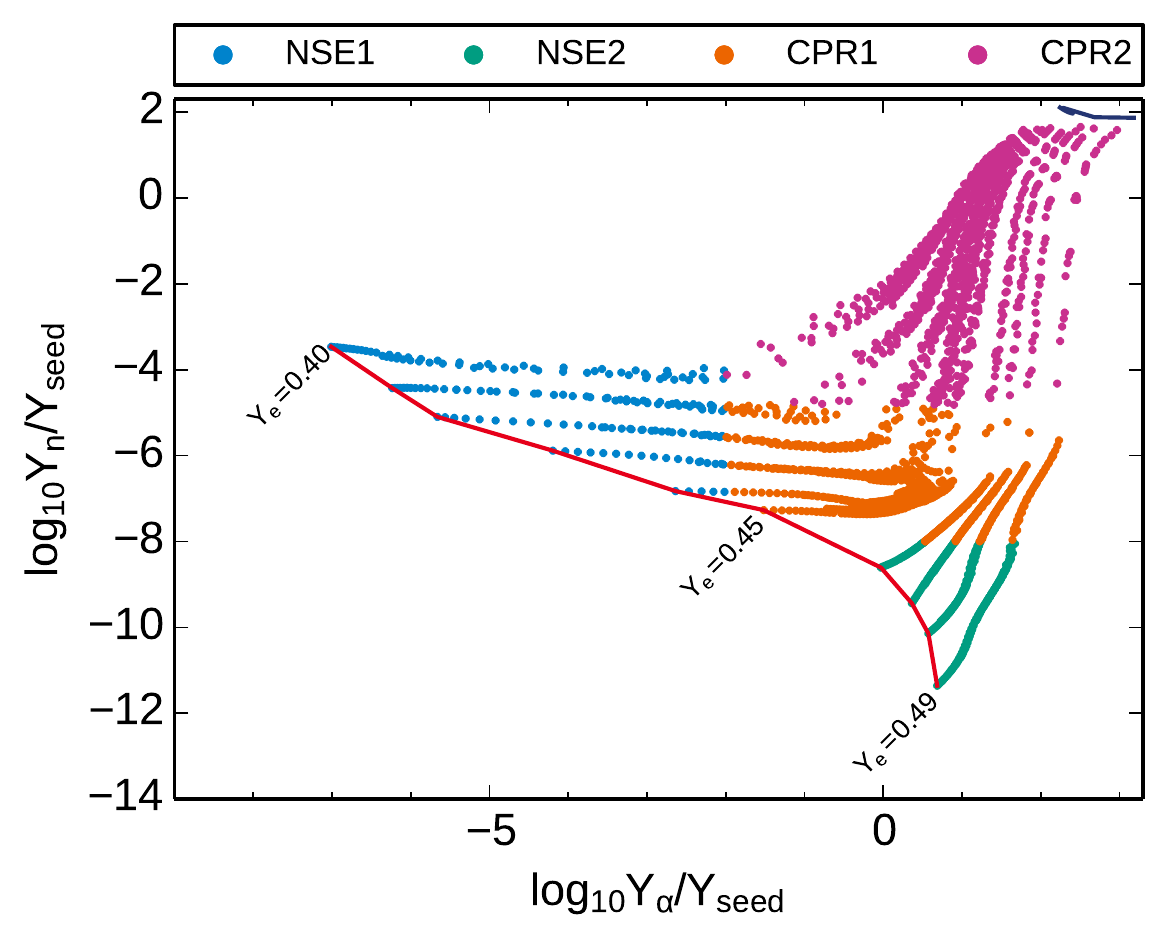}
\caption[]{Different nucleosynthesis patterns in the $Y_{\alpha}/Y_{\mathrm{seed}}-Y_{\mathrm{n}}/Y_{\mathrm{seed}}$ plane. The colors describing the different nucleosynthesis groups are the same as in Sects.~\ref{sec:nse1}--\ref{sec:cpr2}. The red and blue lines mark the constraints of the proto-neutron star masses and radii used in the steady-state model on $Y_{\alpha}/Y_{\mathrm{seed}}$ and $Y_{\mathrm{n}}/Y_{\mathrm{seed}}$. The red (blue) line corresponds to $M_{\mathrm{ns}}=0.8$~$M_{\odot}$ ($M_{\mathrm{ns}}=2.0$~$M_{\odot}$) and $R_{\mathrm{ns}}=30$~km ($R_{\mathrm{ns}}=9$~km). Each chain represents a constant $Y_{\mathrm{e}}$.}
\label{fig:YnYseedvsYalphaYseed}
\end{figure}

For typical supernova conditions, we find four characteristic abundances patterns produced either mainly during the NSE evolution phase or through charged particle reactions (CPR) after NSE. Figure~\ref{fig:nuc_types} gives an overview of elemental abundances at different temperatures together with the final abundances for the different groups. The four nucleosynthesis groups are defined by their $Y_{\mathrm{n}}/Y_{\mathrm{seed}}$ and the $Y_{\alpha}/Y_{\mathrm{seed}}$ at $T \approx 3$~GK, following a similar strategy as in \cite{Wanajo.etal:2017}. These ratios are shown for the different groups in Fig.~\ref{fig:YnYseedvsYalphaYseed}, where every point corresponds to a single trajectory evolution for typical supernova conditions. The red line (limiting the phase space towards low $\ynyseed$ and low $\yayseed$) links those steady-state solutions based on the lowest $M_{\mathrm{ns}}$ and largest $R_{\mathrm{ns}}$. Below this line, there are almost no physical solutions for the wind equations (Eqs.~\ref{eq:ndw1}--\ref{eq:ndw3}). The few physical solutions found are discarded because they are based on massive proto-neutron stars with small radii and thus excluded by causality or, in few cases, they are subsonic breeze solutions. Additional trajectories corresponding to the most compact proto-neutron star (Tab.~\ref{tab:wind_input}) are shown by a blue line (upper, right corner).  The trajectories for the two limiting cases are shown with same colours as in Fig.~\ref{Fig.:Impact_MnsRnsLv_TempDens}.  We have not included possible solutions with $\ynyseed \gtrsim 100$ since such high amount of neutrons is not found in current simulations of standard neutrino-driven supernova explosions. Neither solutions with high $\yayseed$ are shown in the figures, these can be reached by increasing $Y_{\mathrm{e}}$ towards proton-rich conditions.

\begin{figure}[h!]
\centering
\includegraphics[width=1\linewidth]{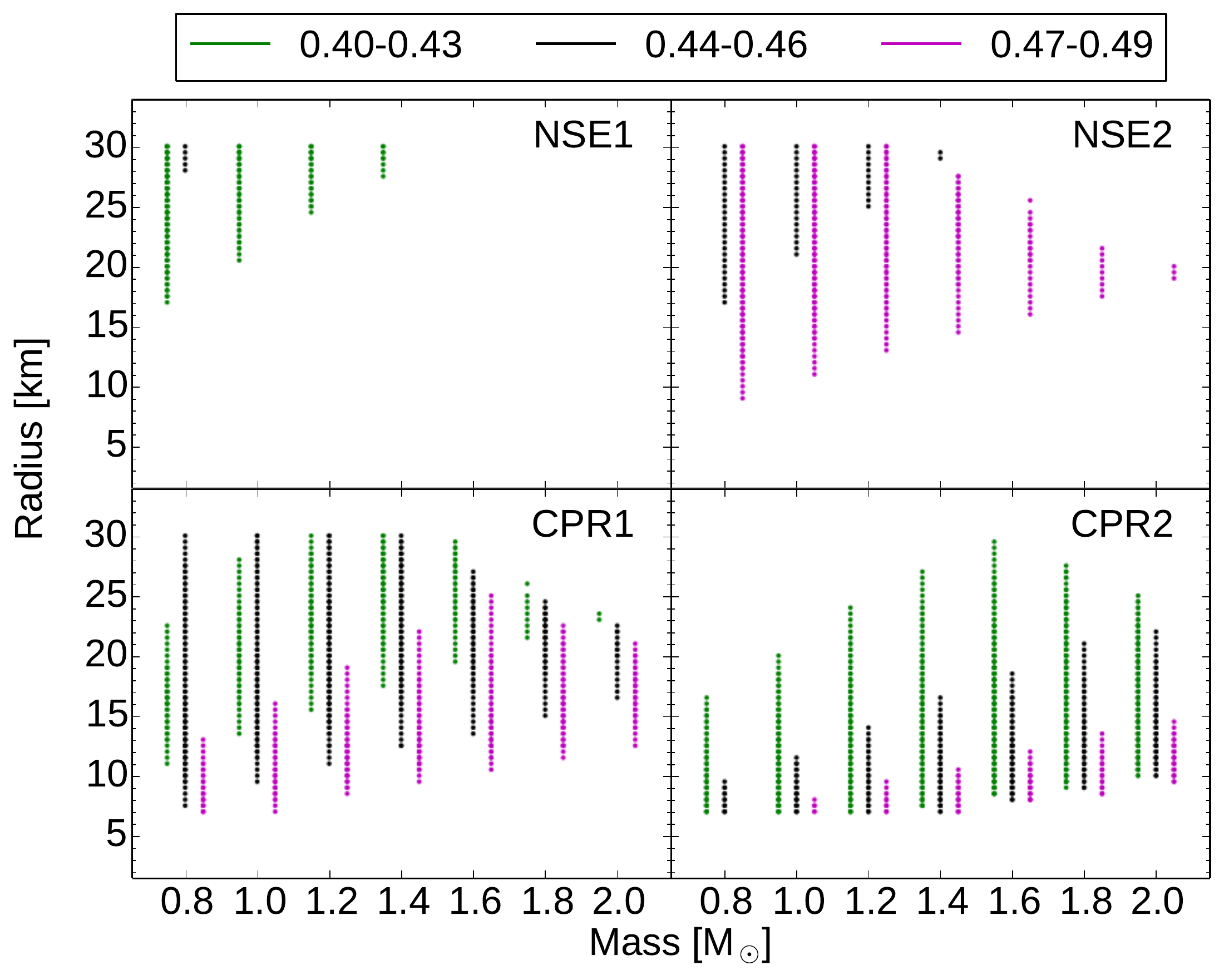}
\caption{Dependencies of the groups NSE1 (upper left panel), NSE1 (upper right panel), CPR1 (lower left panel), and CPR2 (lower right panel) on proto-neutron star mass and radius. The values of the proto-neutron star mass are only the ones on the axis going from 0.8 to 2.0 in intervals of 0.2. The different colors indicate various $Y_{\mathrm{e}}$ ranges.}
\label{fig:MRYe}
\end{figure}

\begin{figure}[h!]
\centering
\includegraphics[width=1\linewidth]{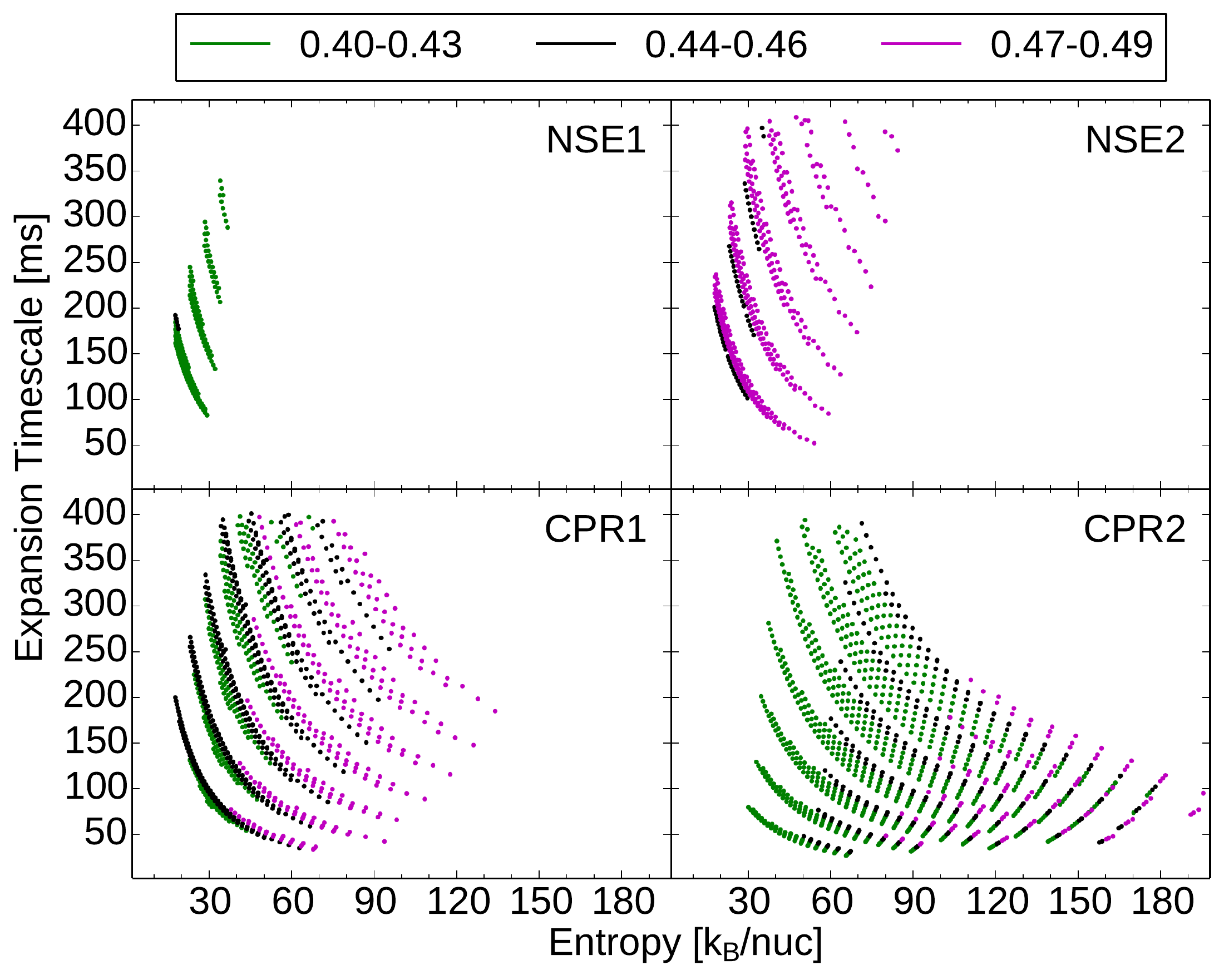}
\caption{Different nucleosynthesis groups depending on entropy and expansion timescale.}
\label{fig:StauYe}
\end{figure}

In the following we describe the nucleosynthesis of every group. In addition to Fig.~\ref{fig:YnYseedvsYalphaYseed}, the dependencies of the groups on proto-neutron star mass and radius, and on entropy and time scale are shown in Fig.~\ref{fig:MRYe} and Fig.~\ref{fig:StauYe}, respectively.  In these figures, every panel corresponds to a nucleosynthesis group and the different colors indicate various ranges of electron fractions. In Fig.~\ref{fig:MRYe} the points from models with different $Y_{\mathrm{e}}$ are shifted to avoid hiding them when overlapping. Proto-neutron star masses shown in intervals of 0.2~$\mathrm{M_{\odot}}$.  

\subsection{NSE1}
\label{sec:nse1}

The trajectories that lead to NSE1 patterns are produced by low mass and large radius proto-neutron stars (Fig.~\ref{fig:MRYe}). Therefore, the proto-neutron stars are not very compact and thus the wind entropy is relatively low (Fig.~\ref{fig:StauYe}). The $Y_{\mathrm{e}}$, with values between 0.40--0.43, is low when comparing to supernova simulations. Still these conditions can mimic some early ejecta that has been exposed to neutrinos only shortly. Moreover, in the early explosion phase the proto-neutron star is still less massive and its radius is large, as it is the case for the trajectories of the group NSE1.

\begin{figure}[h!]
\centering
\includegraphics[width=0.8\linewidth]{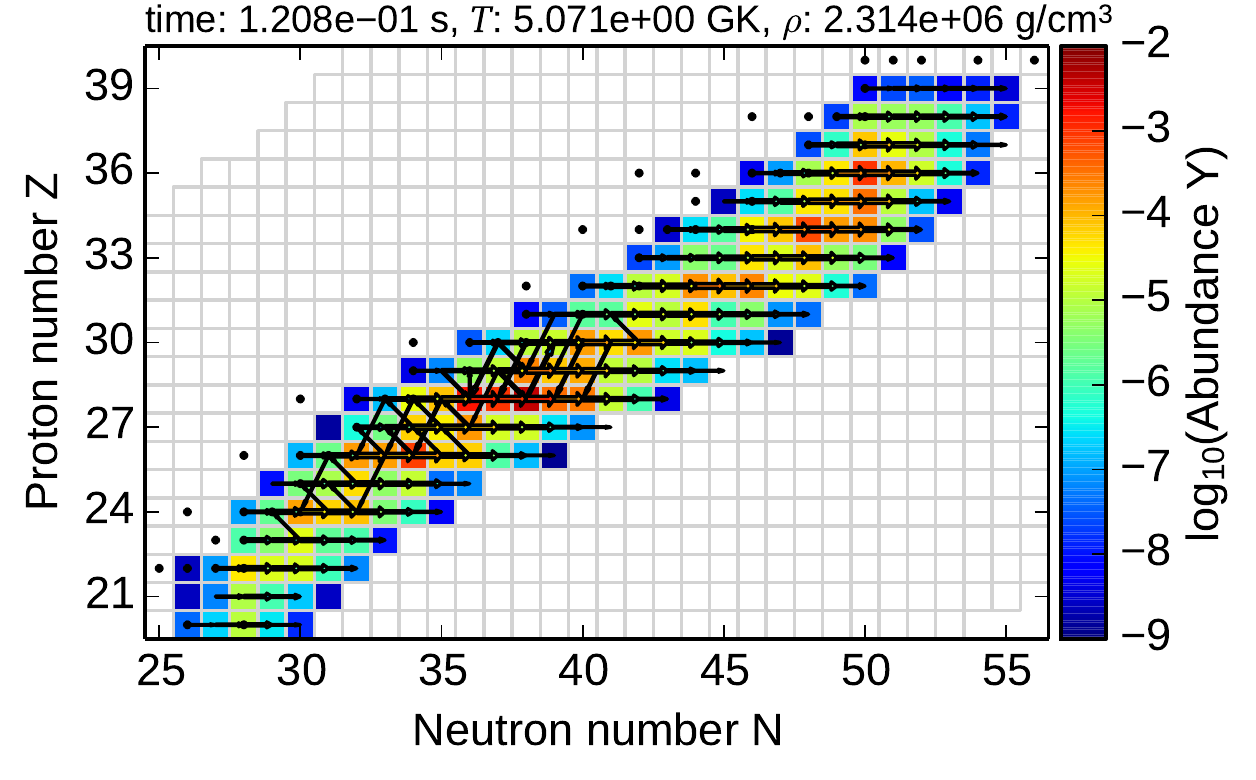} \\
\includegraphics[width=0.8\linewidth]{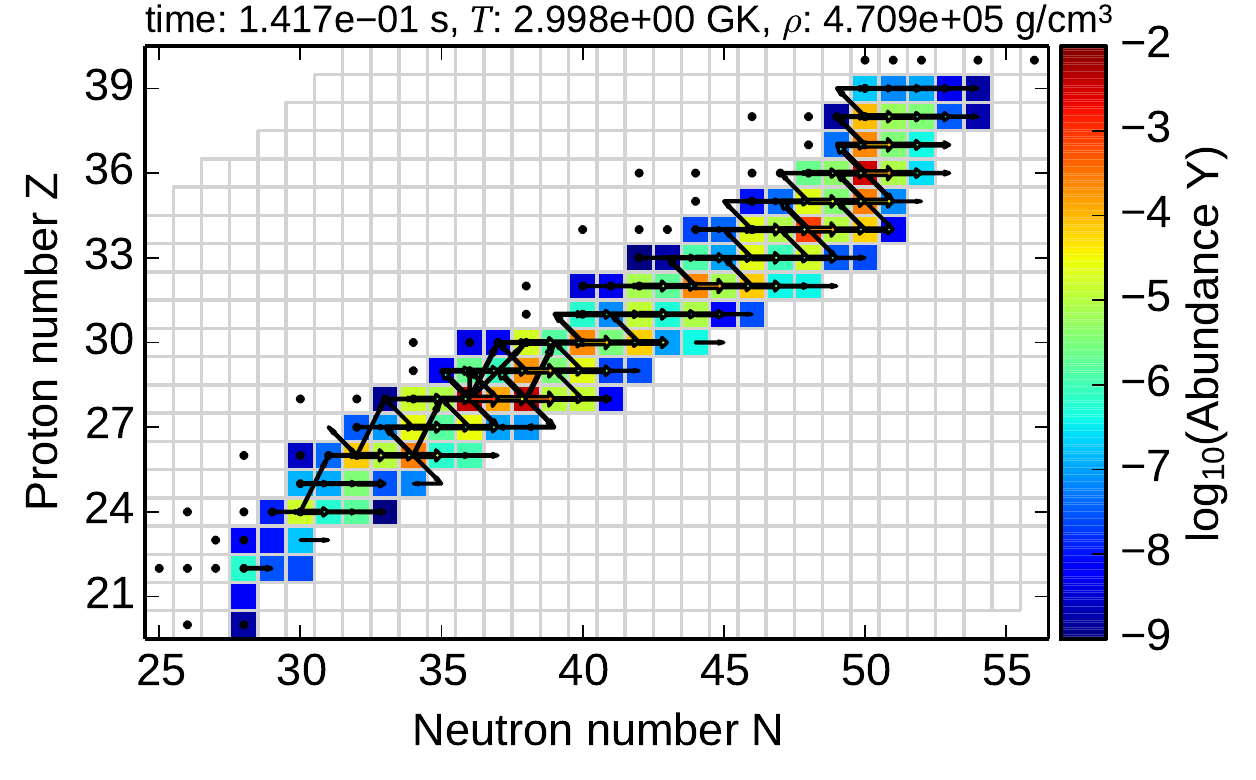} \\
\includegraphics[width=0.8\linewidth]{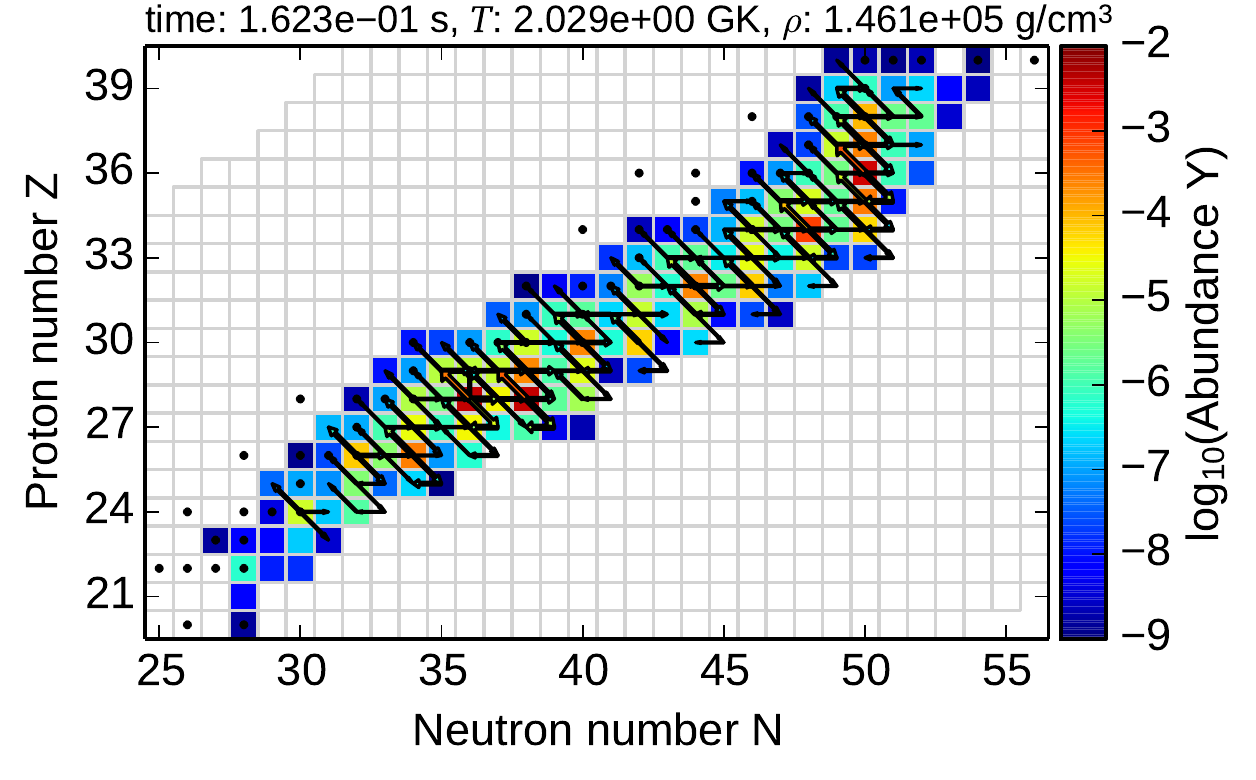}
\caption{Nucleosynthesis evolution of the NSE1 group at $T \approx 5$~GK (top), $T \approx 3$~GK (middle), and $T \approx 2$~GK (bottom). The arrows indicate the flow of the different reactions. The abundances are shown by different colors and stable nuclei are displayed by black dots.}
\label{fig:Flux_NSE1}
\end{figure}

In NSE1, the initial nucleosynthesis evolution is characterized by the sequence of three-body reactions $\alpha(\alpha n,\gamma)^{9}\mathrm{Be}$ and $^{9}\mathrm{Be}(\alpha,\gamma)$, which bypass the 3-$\alpha$ reaction bottleneck \citep{Woosley.Hoffman:1992}. This group is similar to the one identified by \cite{Wanajo.etal:2017} as NSE. Due to the small $\ynyseed$, the nucleosynthesis path evolves near the valley of stability on the neutron-rich side. At $T \approx 6$~GK matter moves along the Ca-Zn region, and reaches $Z \sim 40$ at $T \approx 5$~GK (see Fig.~\ref{fig:Flux_NSE1}, top panel), where the most abundant elements are Fe and Ni (left panel, first row, Fig.~\ref{fig:nuc_types}). Between $T \approx 5-3$~GK, there is only a redistribution of matter by few charged particle reactions, as seen in the middle panel of Fig.~\ref{fig:Flux_NSE1}. The nucleosynthesis path cannot extend beyond the neutron shell closure $N=50$ because of the small amount of free neutrons and alpha particles. The few alpha particles are not sufficient to recombine and fill the abundances for $Z=3-19$ at low temperatures (see first row, Fig.~\ref{fig:nuc_types}). Consequently, the major abundance peaks are already formed around $T \approx 5$~GK at the end of NSE, and the subsequent evolution does not significantly change the abundance pattern. Therefore, the abundance distribution for the  NSE1 group is mainly determined by binding energies and partition functions, and not so much by specific reactions. Finally, during the decay to stability (Fig.~\ref{fig:Flux_NSE1}, bottom panel), the abundance pattern changes slightly. The final abundance pattern (right panel, first row, Fig.~\ref{fig:nuc_types}) exhibits characteristic Ni (not for all trajectories), Zn, and Kr peaks. Elements heavier than $Z \approx 38$ are not synthesized.

\subsection{NSE2}
\label{sec:nse2}

NSE2 patterns are obtained for various compactness of the proto-neutron star, but as in NSE1, the patterns are still dominated by low mass, large radius proto-neutron stars (Fig.~\ref{fig:MRYe}). The range of possible entropies is larger than in NSE1 (Fig.~\ref{fig:StauYe}). However, the main difference is that most of the trajectories have relative high $Y_{\mathrm{e}}$ and this results in very low $\ynyseed$ and high $\yayseed$ (Fig.~\ref{fig:YnYseedvsYalphaYseed}). Under such conditions, the nucleosynthesis path flows through the proton-rich side, as described below.

\begin{figure}[h!]
\centering
\includegraphics[width=0.8\linewidth]{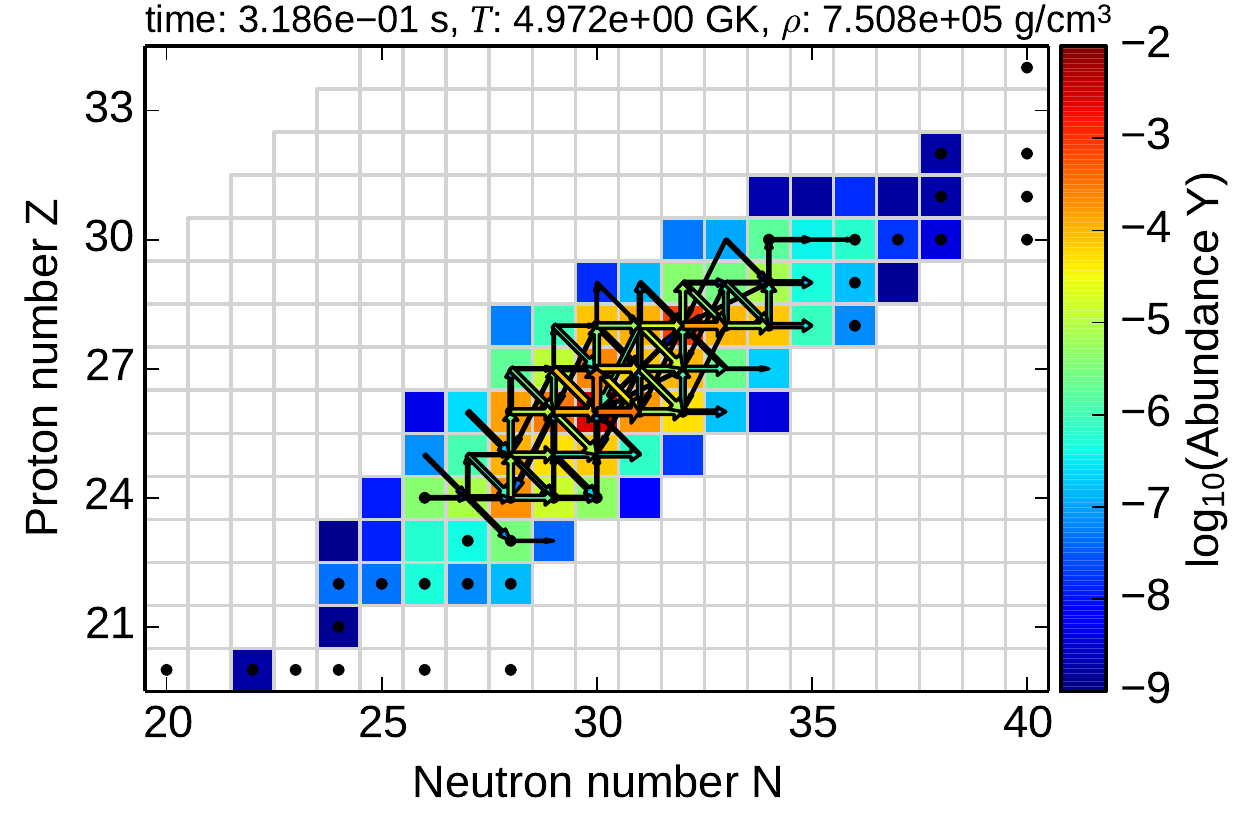} \\
\includegraphics[width=0.8\linewidth]{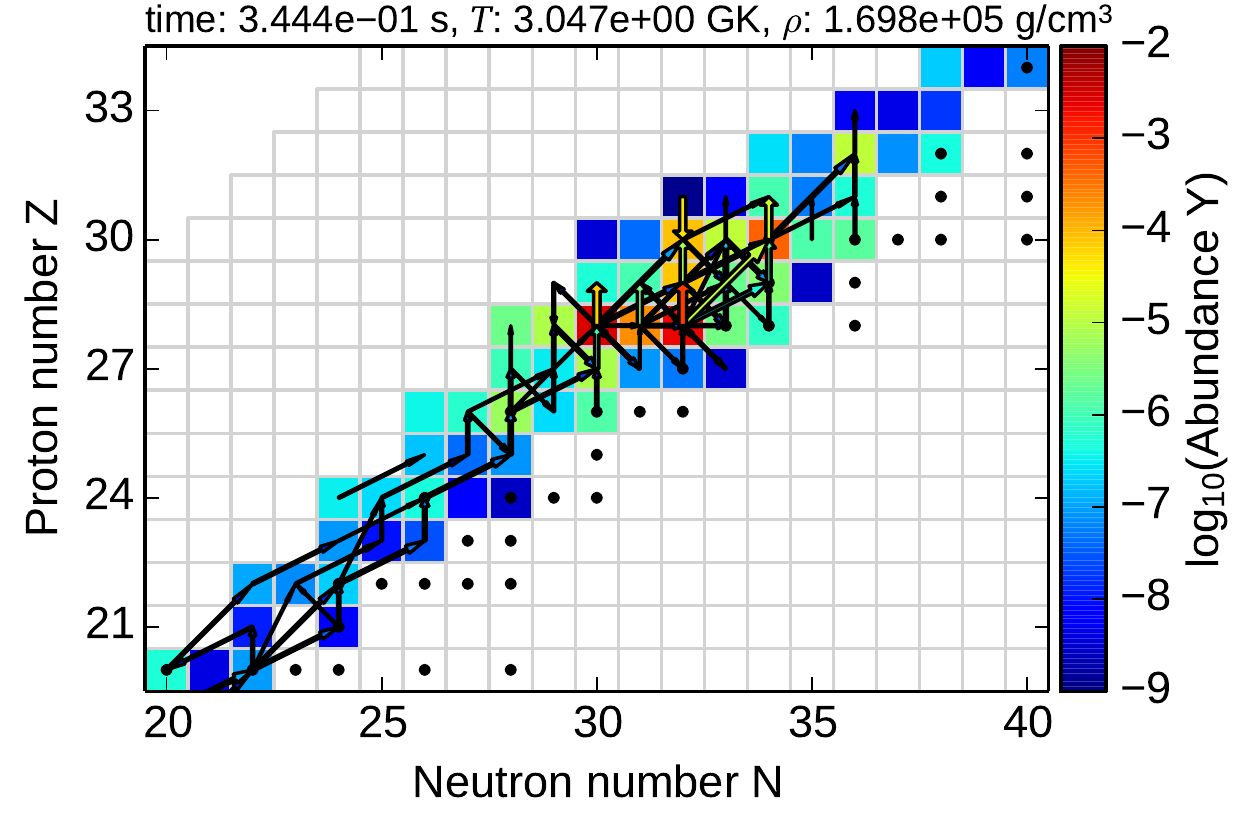} \\
\includegraphics[width=0.8\linewidth]{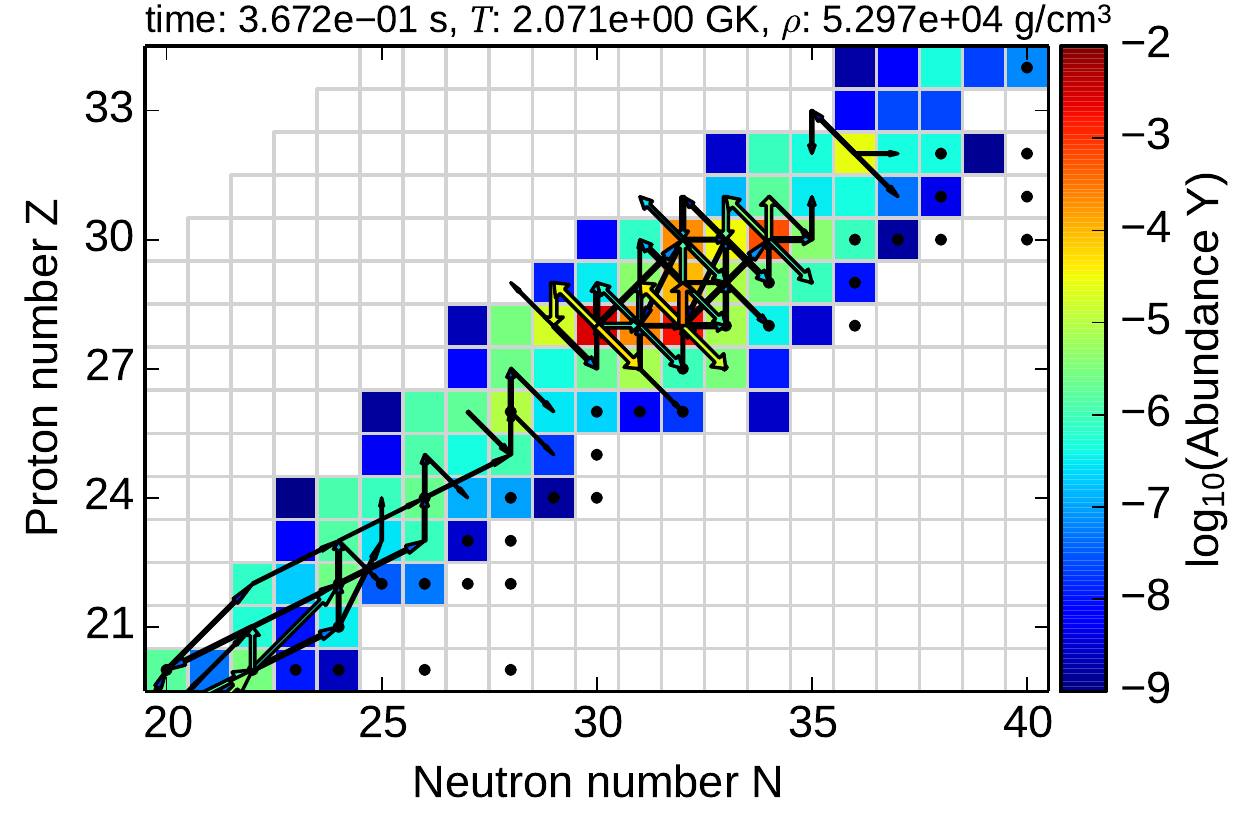}
\caption{Evolution of the abundances of group  NSE2 at $T \approx 5$~GK (top), $T \approx 3$~GK (middle), and $T \approx 2$~GK (bottom).}
\label{fig:Flux_NSE2}
\end{figure}

The final abundance pattern for the NSE2 group exhibits a characteristic peak at $Z=28$ and for some trajectories also at $Z=26$ and/or $Z=30$ (see second row Fig.~\ref{fig:nuc_types}). Elements heavier than $Z=30$ are only formed for $Y_{\mathrm{n}}/Y_{\mathrm{seed}} > 10^{-9}$. In contrast to NSE1, there are some changes as the temperature drops. As shown in Fig.~\ref{fig:nuc_types} (second row, left panel), at $T \approx 5$~GK, matter is accumulated mainly between $Z=22-30$ and the most abundant elements are Fe and Ni. The neutron abundances are very low and thus the nucleosynthesis path moves away from the valley of stability on the proton-rich side via $(\mathrm{p},\gamma)$ and $(\mathrm{p},\mathrm{n})$ reactions (see Fig.~\ref{fig:Flux_NSE2}-top).
For temperatures between $T \approx 4-3$~GK, matter is shifted from Fe to Ni by $(\mathrm{p},\gamma)$, $(\mathrm{p},\mathrm{n})$, and  $(\alpha,\mathrm{p})$ reactions (see Fig.~\ref{fig:Flux_NSE2}, middle panel). Nickel and zinc act as bottlenecks in the nucleosynthesis evolution, and thus are the most abundant elements. When the temperature drops below 2~GK, there is only a redistribution of matter (Fig.~\ref{fig:Flux_NSE2}, bottom panel). 

\subsection{CPR1}
\label{sec:cpr1}

The group CPR1 marks a transition from groups NSE1 or NSE2 to  CPR2 (Fig.~\ref{fig:YnYseedvsYalphaYseed}). In this group, proto-neutron stars can be massive and several trajectories come from small-radius proto-neutron stars (Fig.~\ref{fig:MRYe}). The more compact proto-neutron stars result in higher entropies than in groups NSE1 and NSE2 (Fig.~\ref{fig:StauYe}). 

\begin{figure}[h!]
\centering
\includegraphics[width=0.8\linewidth]{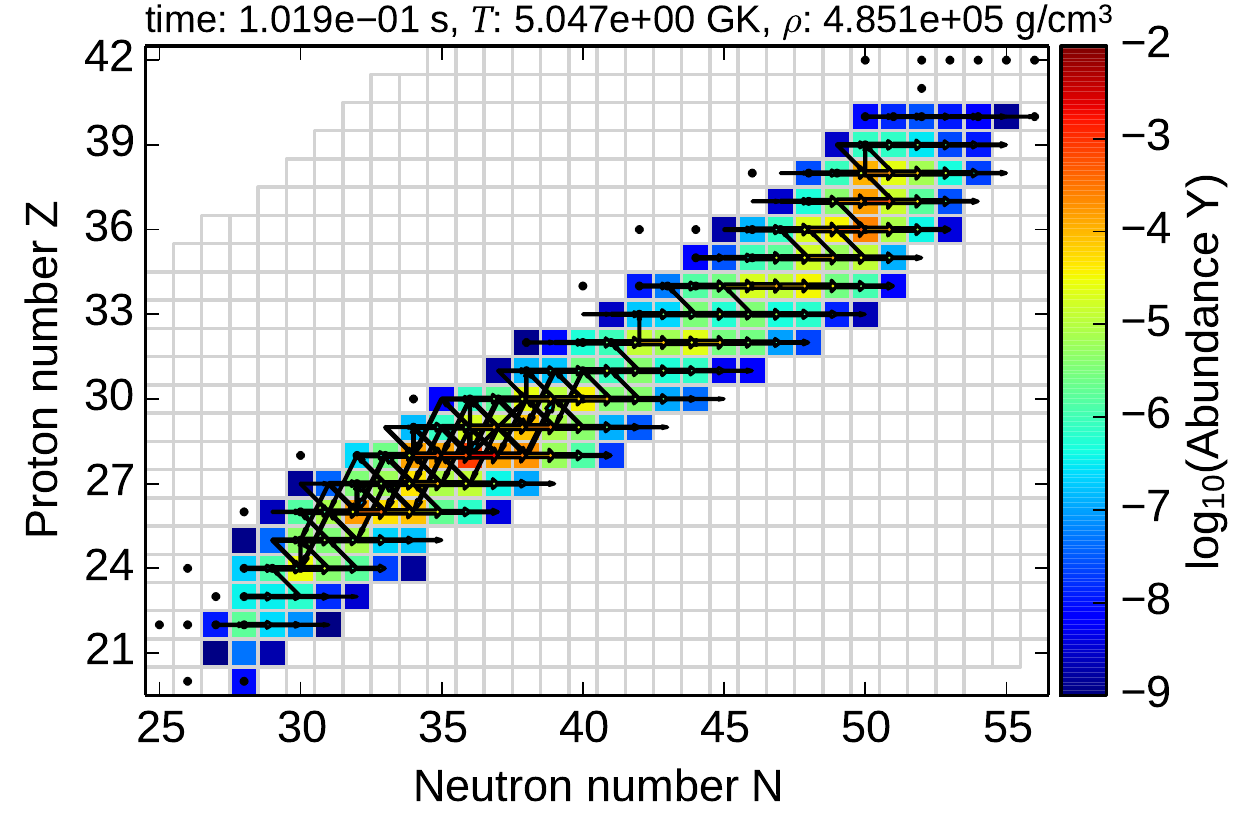} \\
\includegraphics[width=0.8\linewidth]{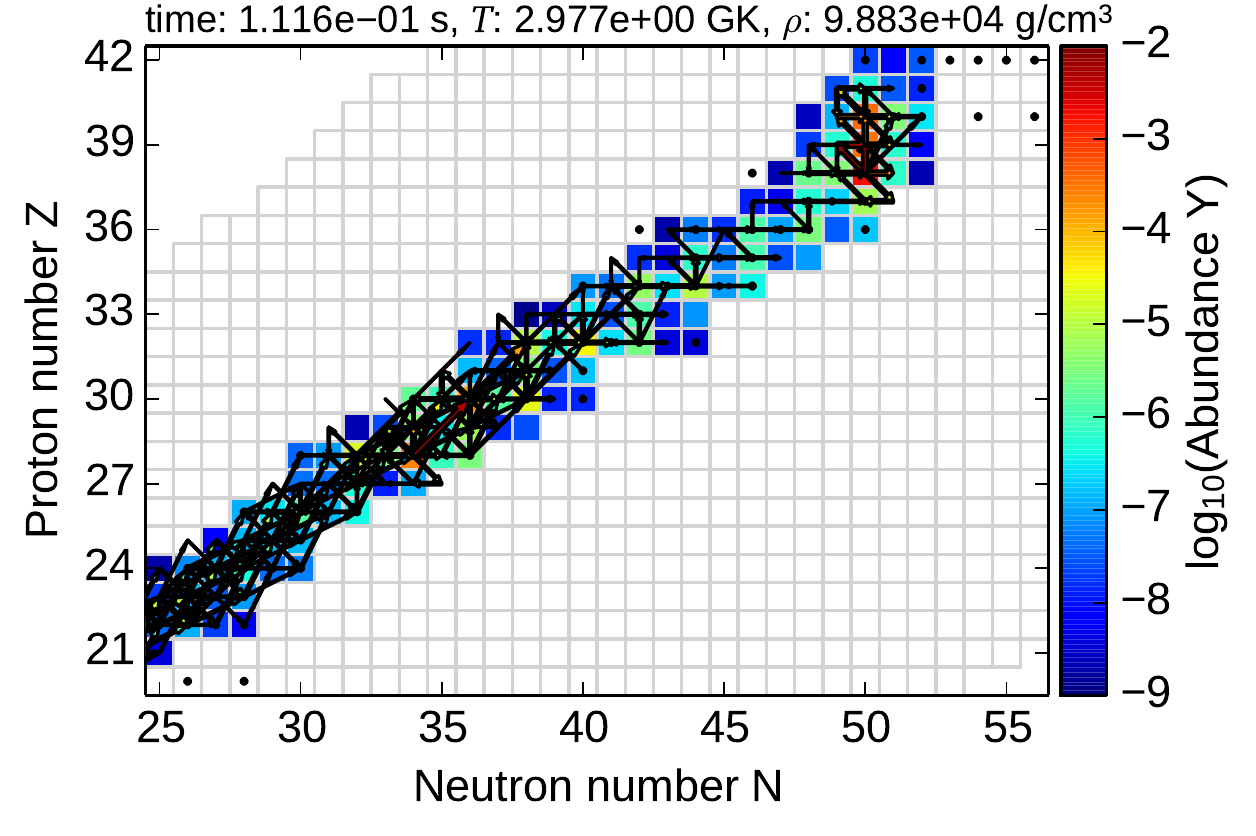} \\
\includegraphics[width=0.8\linewidth]{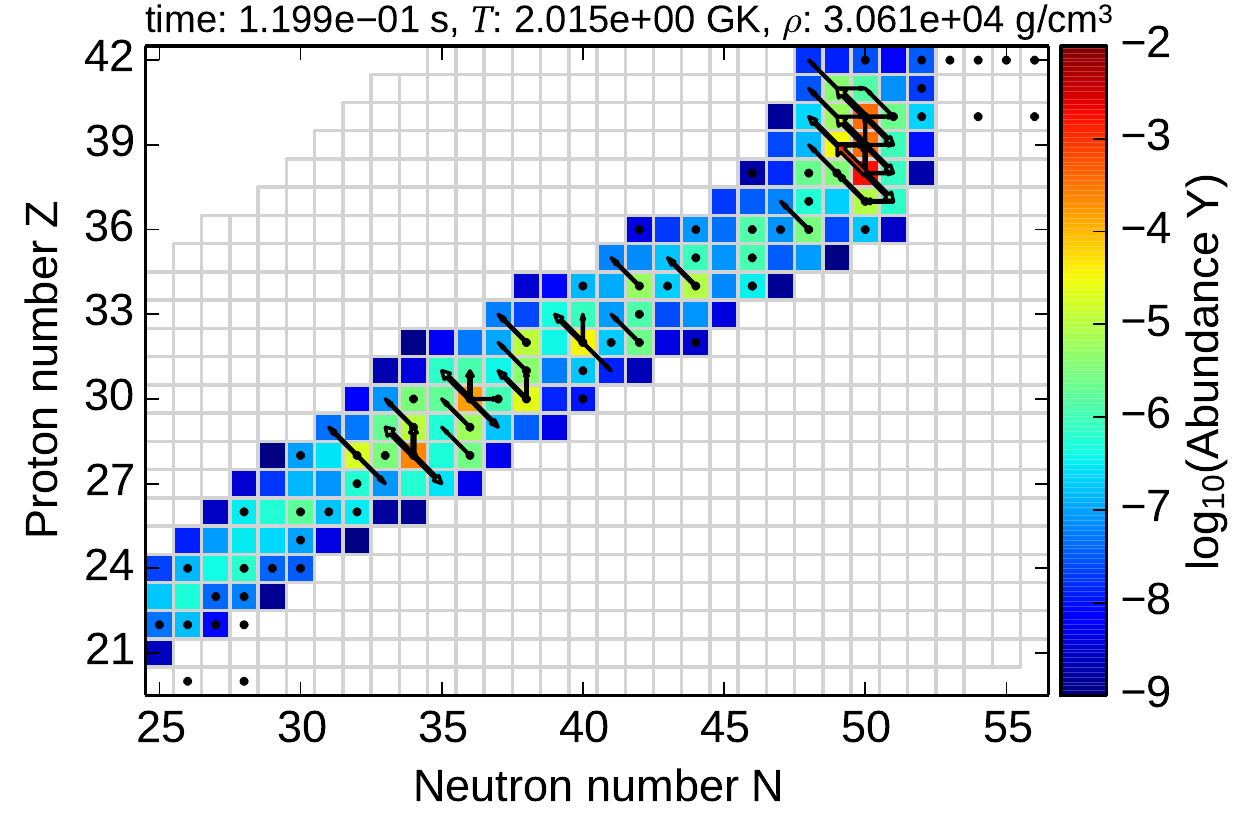}
\caption{Abundance flows of the CPR1 group.}
\label{fig:Flux_CPR1}
\end{figure}

For this group, the abundance evolution and final abundances are shown in the third row in Fig.~\ref{fig:nuc_types}. The nucleosynthesis path proceeds through a series of $(\alpha,\mathrm{n})$ and $(\mathrm{p},\mathrm{n})$ reactions on the neutron-rich side of stability. As temperature drops down from $T \approx 6$~GK to $T \approx 5$~GK, the nucleosynthesis path moves from the Ca-Zn region to nuclei around $Z=39$, with some $(\alpha,\mathrm{n})$ and $(\mathrm{p},\mathrm{n})$ frozen out (see Fig.~\ref{fig:Flux_CPR1}-top). At $T \approx 5$~GK, the most abundant elements are Fe, Ni, and nuclei at $N=50$ (left panel, third row, Fig.~\ref{fig:nuc_types}). When the temperature decreases to $T \approx 4$~GK, matter is redistributed by $(\mathrm{p},\mathrm{n})$ and $(\mathrm{p},\gamma)$ reactions. Most abundant are Fe, Co, Ni, Cu, Zn, and nuclei at $N=50$. At $T=3$~GK, the path stays along stable nuclei and matter has accumulated at $N=50$ (Fig.~\ref{fig:Flux_CPR1}, middle panel) because the alpha abundance is not large enough to overcome the negative Q-values of $(\alpha,\mathrm{n})$ of those nuclei. However, the amount of alphas is still enough to increase the  abundances for $Z=6-20$ via alpha capture reactions (third row, Fig.~\ref{fig:nuc_types}). For lower temperatures, there is only a redistribution of matter and decay to stability (Fig.~\ref{fig:Flux_CPR1}, bottom panel). The overall final abundance pattern has distinctive peaks at Ni, Zn, and Sr (right panel, third row, Fig.~\ref{fig:nuc_types}). For some steady-state trajectories, there is also an abundance peak at Kr. Heavier elements than Zr are not formed due to the small $Y_{\alpha}/Y_{\mathrm{seed}}$ and the negative Q-values of some $(\alpha,\mathrm{n})$ reactions at $N=50$. Thus, the final abundances are mainly determined by the Q-values of $(\alpha,\mathrm{n})$ reactions at $N=50$ (see also \cite{Hoffman.etal:1996,Wanajo:2006}).

\subsection{CPR2}
\label{sec:cpr2}

This is the group with the most extreme astrophysical conditions with some trajectories reaching high entropies (Fig.~\ref{fig:StauYe}) and thus having a relative high $\ynyseed$ (Fig.~\ref{fig:YnYseedvsYalphaYseed}). Most trajectories have small $Y_{\mathrm{e}}$. Therefore, this group is characterised by a nucleosynthesis evolution on the neutron-rich side and the abundances can reach heavier elements than in the other groups. The conditions indicated by Fig.~\ref{fig:MRYe} and Fig.~\ref{fig:StauYe} can be found in some early, neutron-rich ejecta \citep{Wanajo.etal:2011a,Wanajo.etal:2013a,Wanajo.etal:2013c,Wanajo.etal:2017} when the proto-neutron star is still large and not very massive and perhaps also during the wind evolution if the conditions are neutron-rich.

\begin{figure}[h!]
\centering
\includegraphics[width=0.8\linewidth]{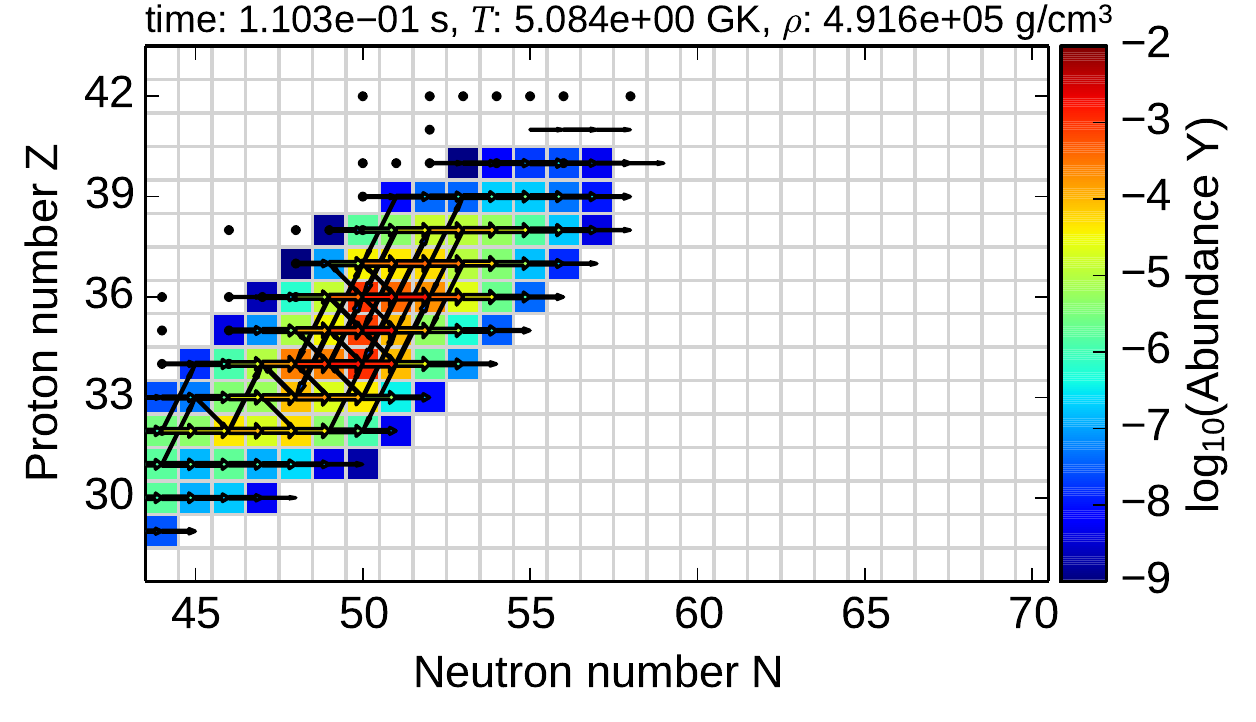} \\
\includegraphics[width=0.8\linewidth]{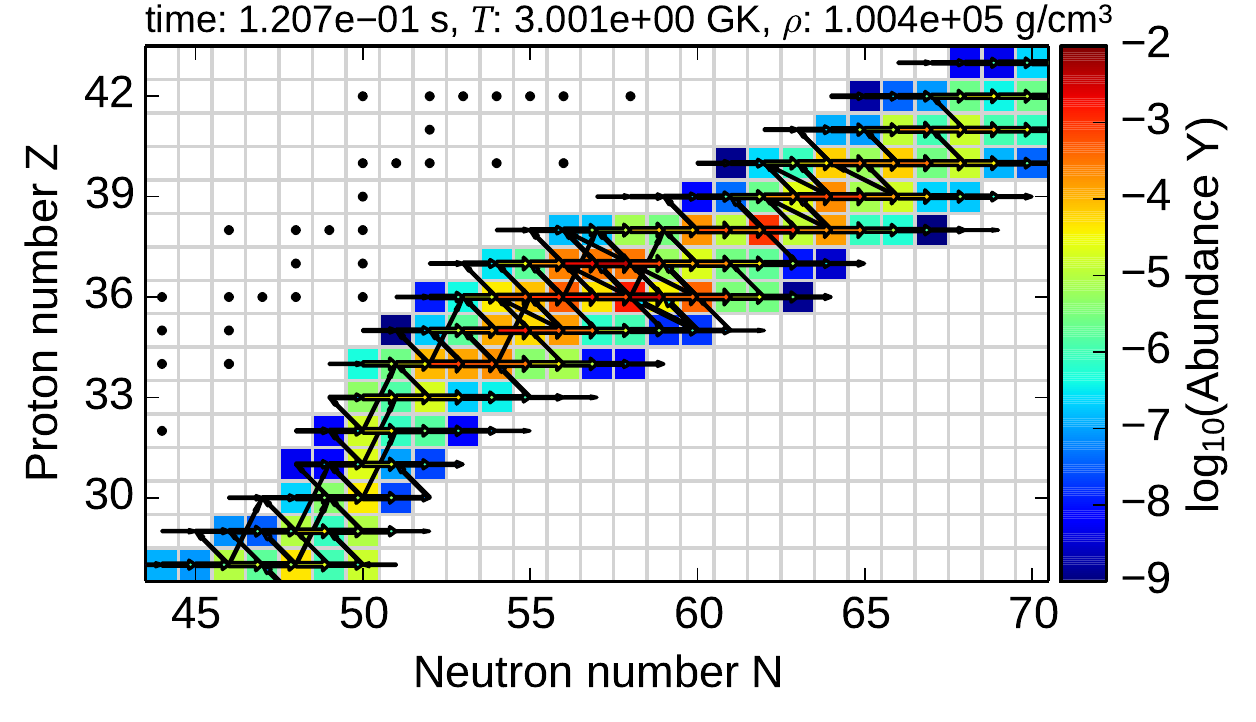} \\
\includegraphics[width=0.8\linewidth]{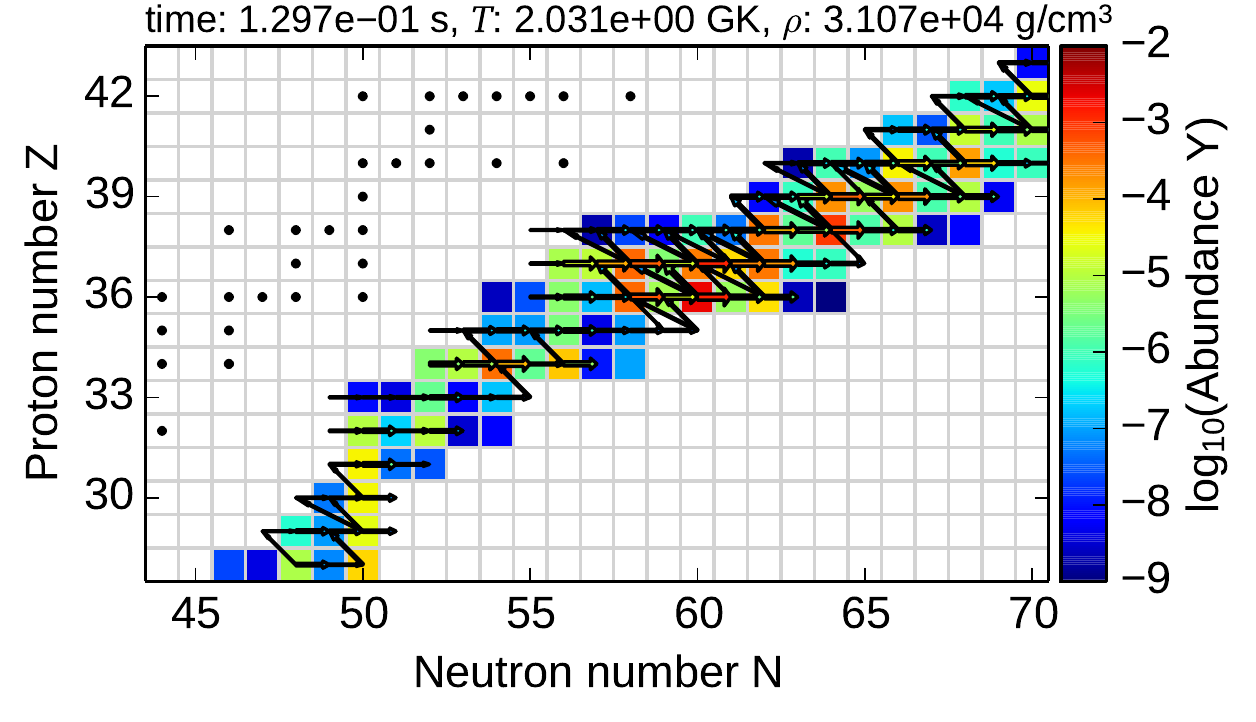}
\caption{Flux diagram for the CPR2 group.}
\label{fig:Flux_CPR2}
\end{figure}

Around $T \approx 6$~GK the nucleosynthesis path proceeds close to stability via alpha capture reactions and especially $(\alpha,\mathrm{n})$ reactions. Most of the matter is accumulated between $Z \approx 20-30$. When the temperature decreases to $T \approx 5$~GK, the path has reached $Z=36$ (bottom row, Fig.~\ref{fig:nuc_types}). The most abundant nuclei are in the neutron shell closure $N=50$, away from the valley of stability (Fig.~\ref{fig:Flux_CPR2}, top panel). At $T \approx 4$~GK, there are no free protons left. Between $T \approx 4-3$~GK, the neutron and alpha abundances are large and the nucleosynthesis flow can overcome the negative Q-value of $(\alpha,\mathrm{n})$ reactions for $N=50$ nuclei, moving matter up to $Z \sim 42$ (Fig.~\ref{fig:Flux_CPR2}, middle panel). The most abundant elements are Kr, Rb, and Sr (see panel for 3~GK, bottom row, Fig.~\ref{fig:nuc_types}). Remarkable are the substantial changes in the overall abundance pattern when the temperature decreases from 5~GK to 3~GK. At $T=2$~GK, the most abundant elements do not change and the abundances are redistributed within isotopic chains (Fig.~\ref{fig:Flux_CPR2}, bottom panel). It is important to mention however, that the abundance pattern for this group varies for different steady-state trajectories (i.e., different $Y_{\mathrm{n}}/Y_{\mathrm{seed}}$ and $Y_{\alpha}/Y_{\mathrm{seed}}$). The overall final abundance pattern exhibits peaks at Kr (differently pronounced for different steady-state trajectories) and Zr. We find various patterns for Kr, Rb, Sr, and Y. In comparison to the other nucleosynthesis groups, heavier elements are synthesized (see bottom row, right panel, Fig.~\ref{fig:nuc_types}). In addition, the heaviest elements vary for different steady-state trajectories, and thus depend on $Y_{\alpha}/Y_{\mathrm{seed}}$ and $Y_{\mathrm{n}}/Y_{\mathrm{seed}}$.

In this group there is more variability of patterns than in other groups. However, the trajectories assigned to this group have in common that the nucleosynthesis evolves beyond $N=50$ and nuclei heavier than $Z \sim 40$ are formed. Moreover, only for the group CPR2, individual reactions, especially $(\alpha,\mathrm{n})$ reactions, play a critical role to determine the abundances which combined with the fact that the reaction rates are rather uncertain \citep{Mohr:2016,Pereira.Montes:2016,Bliss.etal:2017} lead to variations in the final abundances. In a Monte Carlo study (\cite{Blissetal.inprep}) we use representative abundances of group CPR2 to identify the most relevant $(\alpha,\mathrm{n})$ reactions.


\section{Conclusions}
\label{sec:conclusions}

We have systematically studied the neutron-rich neutrino-driven wind based on a steady-state model. We have chosen the input parameters $M_{\mathrm{ns}}$, $R_{\mathrm{ns}}$, and $Y_{\mathrm{e}}$ in agreement with observations and theoretical calculations of neutron stars and supernovae. We have identified four characteristic nucleosynthesis patterns that can be separated by their $\ynyseed$ and $\yayseed$ values once the temperature in the outgoing mass shell has decreased to 3~GK.

The abundance distributions of the NSE1 and NSE2 groups are mainly determined during nuclear statistical equilibrium. The position of the nucleosynthesis path relative to the valley of stability is different between the NSE1 and NSE2 groups. Due to the small $\ynyseed$ and $\yayseed$ the distribution changes only slightly after the breakdown of NSE. Therefore, the final abundances rather depend on binding energies and partition functions than specific reactions. The nucleosynthesis group CPR1 describes the transition from the groups NSE1 or NSE2 to the group CPR2. Charged particle reactions redistribute the abundances after the end of NSE but the $\ynyseed$ and $\yayseed$ are not large enough to overcome the neutron shell closure $N=50$. Thus, the abundances are rather given by Q-values of $(\alpha,\mathrm{n})$ reactions at $N=50$. The abundance patterns within a group are rather similar for different trajectories indicating a comparable nucleosynthesis evolution. This is especially true for groups NSE1, NSE2, and CPR1. In contrast, the abundance distributions (especially the heaviest elements) of group CPR2 vary for different $\ynyseed$ and $\yayseed$. Therefore, individual charged particle reactions can critically influence the abundance evolution.

Our conclusions can be extended to neutrino-driven ejecta even if these are not supersonic. Therefore, this work will help to get an overview of the nucleosynthesis in supernova models without detailed post-processing calculations. Typical trajectories and the corresponding abundances for each group are provided on our web site \url{nuc-astro.eu/} in \texttt{Resources}. These can be used to compare to observations and to explore the impact on the nuclear physics input on the supernova nucleosynthesis.

\section*{Acknowledgments}

J.B., M.W., and A.A. are supported by the Helmholtz-University Young Investigator
grant No. VH-NG-825, Deutsche Forschungsgemeinschaft through SFB 1245, and ERC 677912 EUROPIUM. J.B. thanks the MGK of the SFB 1245 and the JINA Center for the Evolution of the Elements for the research stay at Michigan State University. F.M. and J.P. are supported by Michigan State University and the Facility for Rare Isotope Beams and was funded in part by the NSF under Contracts No. PHY-1102511 (NSCL) and PHY-1430152 (JINA Center for the Evolution of the Elements).

\bibliographystyle{apj.bst}
\bibliography{paper.bib}

\end{document}